\documentclass[amsmath,amssymb,amsfonts,aps,prb,twocolumn,superscriptaddress]{revtex4-1}
\usepackage[utf8]{inputenc}
\usepackage[english]{babel}
\usepackage[T1]{fontenc}
\usepackage{amsmath}
\usepackage{amssymb}
\usepackage{xcolor}
\usepackage{tikz}
\usepackage[caption = false]{subfig}
\usepackage{graphicx}

\usepackage{comment}
\usepackage{braket}

\begin{document}

\title{Error estimation in current noisy quantum computers}


\author{Unai Aseguinolaza}
\affiliation{Basic Sciences Department, Faculty of Engineering, Mondragon Unibertsitatea, 20500 Arrasate, Spain}
\author{Nahual Sobrino}
\affiliation{Donostia International Physics Center, Paseo Manuel de Lardizabal 4, E-20018 San
Sebasti\'an, Spain}
\affiliation{Nano-Bio Spectroscopy Group and European Theoretical Spectroscopy Facility
(ETSF), Departamento de Pol\'imeros y Materiales Avanzados: F\'isica, Qu\'imica y
Tecnolog\'ia, Universidad del Pa\'is Vasco UPV/EHU, Avenida de Tolosa 72, E-20018
San Sebasti\'an, Spain}
\author{Gabriel Sobrino}
\affiliation{aQuantum Software Engineering, Spain}
\author{Joaquim Jornet-Somoza}
\affiliation{Servicios Generales a la Investigación (SGIker), Universidad del Pa\'is Vasco UPV/EHU, Avenida de Tolosa 72, E-20018
San Sebasti\'an, Spain}
\author{Juan Borge}
\email{jborge@mondragon.edu}
\affiliation{Basic Sciences Department, Faculty of Engineering, Mondragon Unibertsitatea, 20500 Arrasate, Spain}

\begin{abstract}
One of the main important features of the noisy intermediate-scale quantum (NISQ) era is the correct evaluation and consideration of errors. In this paper, we analyze the main sources of errors in current (IBM) quantum computers and we present a useful tool  (TED-qc) designed to facilitate the total error probability expected for any quantum circuit. We propose this total error probability as the best way to estimate  a lower bound for the fidelity in the NISQ era,  avoiding the necessity of comparing the quantum calculations with any classical one. 
 In order to contrast the robustness of our tool we compute the total error probability that may occur in three different quantum models: 1) the Ising model, 2) the Quantum-Phase Estimation (QPE), and 3) the Grover's algorithm.
 For each model, the main quantities of interest are computed and benchmarked against the reference simulator's results as a function of the error probability for a representative and statistically significant sample size. The analysis is satisfactory in more than the $99\%$ of the cases. In addition, we study how error mitigation techniques are able to eliminate the noise induced during the measurement. These results have been calculated for the IBM quantum computers, but both the tool and the analysis can be easily extended to any other quantum computer.

\end{abstract}


\maketitle

\section{Introduction}
Quantum computing, i.e. the possibility to access real quantum states to realize complex calculations, has passed from a possibility to a reality \cite{Nielsen2011}. Feynman's idea \cite{feynman1982} of using real quantum systems to simulate quantum mechanics is nowadays not a dream or an idea anymore. In the last 20 years, the capabilities of state-of-the-art quantum computers have improved a lot. As an example, this year IBM was able to implement the first 433 qubit computer\footnote{IBM Unveils 400 Qubit-Plus Quantum Processor and Next-Generation IBM Quantum System Two, https://newsroom.ibm.com/2022-11-09-IBM-Unveils-400-Qubit-Plus-Quantum-Processor-and-Next-Generation-IBM-Quantum-System-Two} and it has foreseen to present a computer with more than 1000 qubits in the near future\footnote{ IBM promises 1000-qubit quantum computer—a milestone—by 2023, https://www.science.org/content/article/ibm-promises-1000-qubit-quantum-computer-milestone-2023}.
Ideal quantum computers are supposed to be able to realize calculations not possible for classical computers with great accuracy. To do so, these computers require at least thousands of qubits in order to use many of them for quantum error corrections \cite{lidar_brun_2013,TerhalRMP2015,Wendin_2017}. Unfortunately, we are still far from this situation, and in the noisy intermediate-scale quantum (NISQ) era, the scientific and technological efforts focus on evaluation, control, and reduction of the physical errors  \cite{BhartiRMP2022,Leymann_2020,Porter2022,Kandala2017,Aspuru-Guzik-Mol-Science2005,Cerezo2021,Arute2019,Preskill2018,Xiao2021,Dalzell2020,georgopoulos2021modeling,patel2020experimental,nation2021scalable,Weidenfeller2022,SetiawanPRX2021,WuPRL2021,HeadleyPRA2022}.

The NISQ devices are composed of a couple of tens of qubits, and presumably a couple of hundreds in the near future. One of their most important characteristics is their imperfect nature, as their name indicates they are noisy. Nowadays, the best-performing quantum computers are based on transmon qubits \cite{transmon}, however, they suffer from four main limitations.  First of all, in these devices, the qubit's state stability is known to be of the order of hundreds of microseconds, so we can not perform very long calculations. In the second order, the gates acting on the qubits are noisy, so every time we perform any calculation we are losing accuracy. In third place, the measurement process of the qubit state is subject to errors. The last thing to take into account is the limited number of available physical qubits.

Normally, the errors induced by these limitations are analyzed by comparing the noisy quantum calculations against the classical noiseless results. This method is really precise but it cannot be used for calculations that exceed the capacity of classical computers. As quantum supremacy is expected in the NISQ era we propose to use the total error probability as the best way to measure the role of errors in this period. 
Our main purpose in this paper is to analyze these limitations by considering representative statistical samples in different quantum algorithms as a function of the total  error probability. By doing it, we will be able to ensure that this total  error probability obtained before any quantum calculation correctly represents an upper bound for the error induced in the actual calculation. Several approaches are described in the literature to estimate the circuit error of a quantum program, which either need the use of a quantum devices \cite{proctor2022,cross2019} or just take into account the gate operation errors \cite{Shin2020,quetschlich2023,vadali2023}. Our method, not only introduce the errors induced by the instability of the qubits, which, as we will see is one of the main sources of the total error, but it can be used to set properly the qubit map before executing in a real quantum device.  

To do so, we first produce a tool that connects with IBM's application programming interface (API) to calculate the total error probability expected during the run of a given quantum circuit. 
Then, we estimate the effect of the total error probability by studying  three different and representative quantum algorithms: the Ising model, the quantum phase estimation (QPE), and the Grover's algorithm. In addition, the fidelity which provides a good measure of similarity between the ideal and real quantum states/calculations. In order to contrast the difference between the ideal result (simulator) and the noisy one (physical quantum computer) we calculate the  magnetization for the Ising model and phase for the QPE, which are the most important outputs of these models.

The effect of the error mitigation technique developed in Qiskit  \cite{abraham2019qiskit} is also evaluated for all these cases. The error mitigation techniques are post-processing routines that try to minimize the errors occurring in quantum computers  \cite{Kandala2019,Berg2022,Mitigation-PRL-2017,Czarnik2021errormitigation,Cai2021,LaRose2022,Suchsland2021,FunckePRA2022}. The one developed in Qiskit consists in the elimination of the error induced during the measurement processes of the physical qubits. Although this technique may succeed in the elimination of measurement errors, it is important to notice that it scales exponentially with the number of qubits, so there might be situations in which its use is computationally really demanding.

The paper is structured as follows, in Sec. \ref{algorithm} we describe the proposed tool to calculate the total error probability.  In Sec. \ref{results}  and Sec. \ref{fidelity_sec} we analyze the  main quantities of interest and the fidelity (respectively) of the three quantum circuits for six different IBM computers and for different qubit chains. In Sec. \ref{mitigation} we comment on the error mitigation routine of Qiskit. Finally, the conclusions are exposed in Sec. \ref{conclusions}.

\section{Calculating the total error probability}
\label{algorithm}
As we pointed out in the previous Section, the estimation of errors in the NISQ era is essential in the field of quantum computing. These errors come basically from three different sources  \cite{Leymann_2020}:

- The  error induced by the instability of the qubits. In order to have non-trivial states we have to excite our physical qubits. The probability of finding the qubit in the excited state decays exponentially with time  \cite{abraham2019qiskit,mckay2018qiskit}, so the larger the quantum circuit the less likely is to find the qubit in the expected state. There are two possible decaying mechanisms, the decay of an excited state to the ground state, i.e. the probability of a state $\ket{1}$ to decay to a $\ket{0}$, and the change of the phase of an excited state, for example passing from the state $1/\sqrt{2}(|0\rangle+|1\rangle)$ to the state $1/\sqrt{2}(|0\rangle-|1\rangle)$. 

-Through the gates applied in the quantum circuit. The evaluation of any quantum gate carries an error with it. Single qubit gates usually carry an error probability  of the order of $10^{-4}$ to $10^{-3}$ while two-qubit gates usually carry an error of the order of $10^{-3}$ to $10^{-2}$.

- Each qubit measurement induces an error due to the lack of precision in the physical act of measure which carries an  error probability  of the order of $10^{-2}$.

\begin{figure*}
  \includegraphics[width=\textwidth,height=4cm]{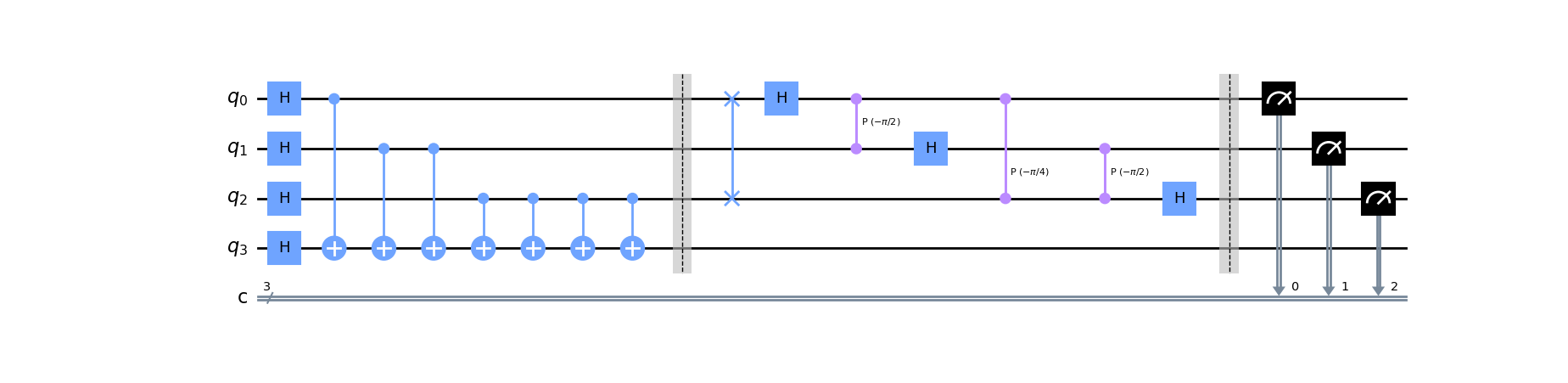}  \includegraphics[width=\textwidth,height=2.5cm]{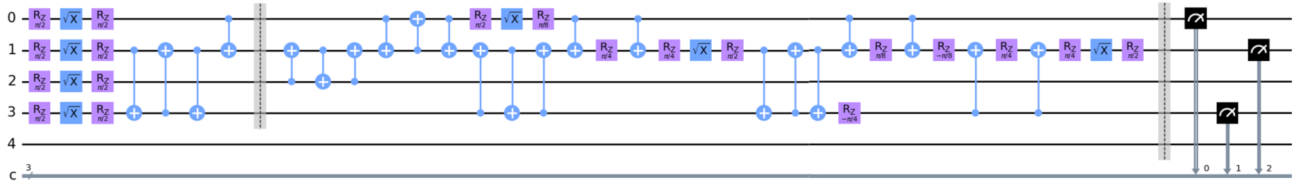}
  \caption{QPE $\sigma_x$ circuit for the $|+\rangle$ state. In the upper figure, the circuit is described in the chosen set of gates. In the lower figure, the circuit is written in the (universal) set of gates implemented in imbq$\_$belem.}
  \label{circuit}
\end{figure*}

All the errors depend on the physical hardware where the circuit is run. 
They depend not only on the specific machine, but also on the specific qubits and connections that are used. For this reason, we have developed a pre-processing computational tool that facilitates the total error probability expected for a given quantum circuit on an IBM quantum machine. We named the project "Tool for Error Description in quantum circuits" (TED-qc).
The code is open-source and available at the GitLab repository (https://gitlab.com/qjornet/ted-qc.git).

We have written our quantum circuits in Qiskit making use of more than one set of universal quantum gates. Once the quantum circuit is written on the chosen basis, it has to be sent to a particular quantum computer. This quantum computer transforms the provided gates to the universal set of quantum gates it operates. In Fig. (\ref{circuit}) we can see how the ibmq$\_$belem transforms the QPE operator $\sigma_x$   for the $|+\rangle$ state circuit described in the qiskit's basis into its own gates. 

All the information needed to calculate the total  error probability can be extracted through IBM's API. First of all, it gives us each gate error probability $P^{\text{gate}}$, which takes into account which qubits have been used.  The same quantum gate has different error probability depending on the qubit that is acting on. It also gives us the error probability committed during the measurement $P^{\text{meas}}$. We will see that this error can be treated and completely reduced using the error mitigation technique. Finally, IBM's API also  provides the relaxation ($\tau_1$) and dephasing ($\tau_2$) times of  the working qubits.  

In order to calculate the total error probability we define the success probability, i.e. the possibility of non-committing any error, $S_{T}$, as
\begin{equation}
	\label{St}
	S_{T}=\prod_{i=1}^{m}S_i=\prod_{i=1}^{m} [1-P_{i}],    
\end{equation}
where $P_{i}$ represents the error probability from any source: single and double qubit gates, measurement, and qubit instabilities, and $m$ is the total number of error sources. The relation between the error probability and success probability is defined as $P+S=1$. The total error probability is therefore computed as
\begin{equation}
	P_T= 1- \prod_{i=1}^{\text{gates}}[1-P_i^{\text{gate}}]\prod_{\alpha=1}^{\text{qbits}} [1- P_{\alpha}^{\text{meas}}][1-P_\alpha^{\tau_1}][1-P_\alpha^{\tau_2}].
	\label{eq_S}
\end{equation}

The first product in Eq.~(\ref{eq_S}) accounts for all gate-related errors $P_i^{\text{gate}}$ (both single and double), whereas the second product encompasses all explicit qubit-dependent sources of error. These include the measurement error, denoted as $P_{\alpha}^{\text{meas}}$, and the errors arising from qubit instability, represented by $P_{\alpha}^{\tau_1}$ and $P_{\alpha}^{\tau_2}$. while the gate and measurement errors are directly obtained through the IBM'S API, to account for errors caused by instability qubits $P_{\alpha}^{\tau_i}$, we consider an exponential decay of the form \cite{abraham2019qiskit}
\begin{equation}
	P_{\alpha}^{\tau_i}=1-e^{-\frac{t_{\alpha}}{\tau_{i,\alpha}}}
    \label{eq_P_tau}
\end{equation}

where $i=1,2$ are the two instability mechanism explained before, $t_{\alpha}$ is the total circuit time of qubit $\alpha$ and $\tau_{i,\alpha}$ are the relaxation and dephasing times of qubit $\alpha$. The evaluation of Eq.~(\ref{eq_P_tau}) requires the knowledge of the time it takes for any qubit between the initialization of the circuit and the measurement $t_{\alpha}$. This is done by adding up how long it takes for qubit $\alpha$ to perform any single gate.
The situation gets more complex when dealing with two-qubit gates. The process of executing a two-qubit gate necessitate the completion of all preceding gate operations on both qubits before the joint operation can be executed. Consequently, after incorporating the time needed for the two-qubit gate into the total time calculation, we're faced with the task of comparing the cumulative times of the two qubits involved.
The crucial point here is that we don't simply add these times together. Instead, we analyze the total times for each qubit and then adopt the longer of the two as the updated qubit time. This approach ensures that we're accounting for the maximum time it could take, thus capturing the full potential for instability-induced error.





\section{The error probability in three representative quantum circuits}
\label{results}

The tool we have developed computes the total error probability of any quantum circuit for any physical qubit chain.
In order to see if the computed total error probability corresponds to the real error induced by the physical qubits, we will perform  calculations in three representative quantum circuits: the one-dimensional Ising model, the QPE for the $\sigma_x$ operator, and the Grover's algorithm, in many different qubit chains and several IBM quantum computers. 

Before evaluating the effect of the total  error probability  in our results it is important to remind how a quantum computer works. Any time we send a job to a quantum computer it makes an important number of repetitions of the same quantum circuit, given by the number of shots, and it extracts the average between all these repetitions. So, if we say that the total error probability is, for example, of the 20$\%$, we are saying that 80$\%$ of the repetitions will give the correct result, but 20$\%$ may be wrong, so the final result will be the linear combination of the 80$\%$ correct wave functions and of the 20$\%$ possibly wrong ones.

In the three mentioned  circuits we will compare the "noisy" results, i.e. the ones obtained in a real quantum computer, with the ones obtained in the simulator.
We will compare the magnetization for the Ising model, the phase for the QPE, and the probability of finding the target number for the Grover algorithm.

We will now comment on the results concerning the three different quantum circuits.

\subsection{The one-dimensional Ising model}
\label{ising}
The Ising model is one of the most studied models in Physics. It explains ferro and antiferromagnetism, but it is also used to describe strongly correlated systems. The Ising model is a great example of the many advantages of quantum computing, as any electron spin  can be easily mapped with a qubit, reducing a $2^n$ problem to a linear one.

Our aim is to diagonalize the one-dimensional Ising Hamiltonian for a $n=4$ antiferromagnetic interaction in the presence of an external magnetic field through the unitary transformation $U$  \cite{CerveraIsing2018,LaTorrePRA2009}.
\begin{equation}
    H=UH_d U^{\dagger},
\end{equation}
where $H_d$ is the diagonalized Hamiltonian, and $H$ the Ising Hamiltonian that reads
\begin{equation}
\label{hamiltonian}
H=\sum_{i=1}^{n-1} \sigma_i^x\sigma_{i+1}^x+\lambda\sum_{i=1}^n\sigma_z .   
\end{equation}
 If we apply the unitary transformation $U$ to the eigenstates of the diagonalized Hamiltonian we will obtain the eigenstates in our original Hamiltonian  basis
\begin{equation}
    |\psi>=U|\psi_d>.
\end{equation}
The details of the construction of $U$ are supplied in  \cite{CerveraIsing2018,LaTorrePRA2009}.

We will calculate the ground state in the Hamiltonian basis for the case of a large external magnetic field, 2.5 times larger than the antiferromagnetic exchange field ($\lambda=2.5$), and later on, we will calculate the magnetization for this ground state. The magnetization is just the difference between spin ups and downs, or in this case between zeros and ones.

We have chosen a big external magnetic field in order to have a ground state which induces a magnetization close to the maximum solution, therefore, our ground state will be similar to the state $|1111\rangle$. Due to this reason,  any possible type of error in the calculation of the circuit will  pop up with a high probability in the calculation of the magnetization. If we would have chosen other values for the external magnetic field closer to the antiferromagnetic exchange field, or even smaller, these would induce  magnetization values around 0  and some of the errors could compensate with the others (the $|0\rangle$ states which become $|1\rangle$ may be compensated by the $|1\rangle$  states becoming $|0\rangle$). 

\begin{figure}
  \centering
  \includegraphics[width=0.45\textwidth]{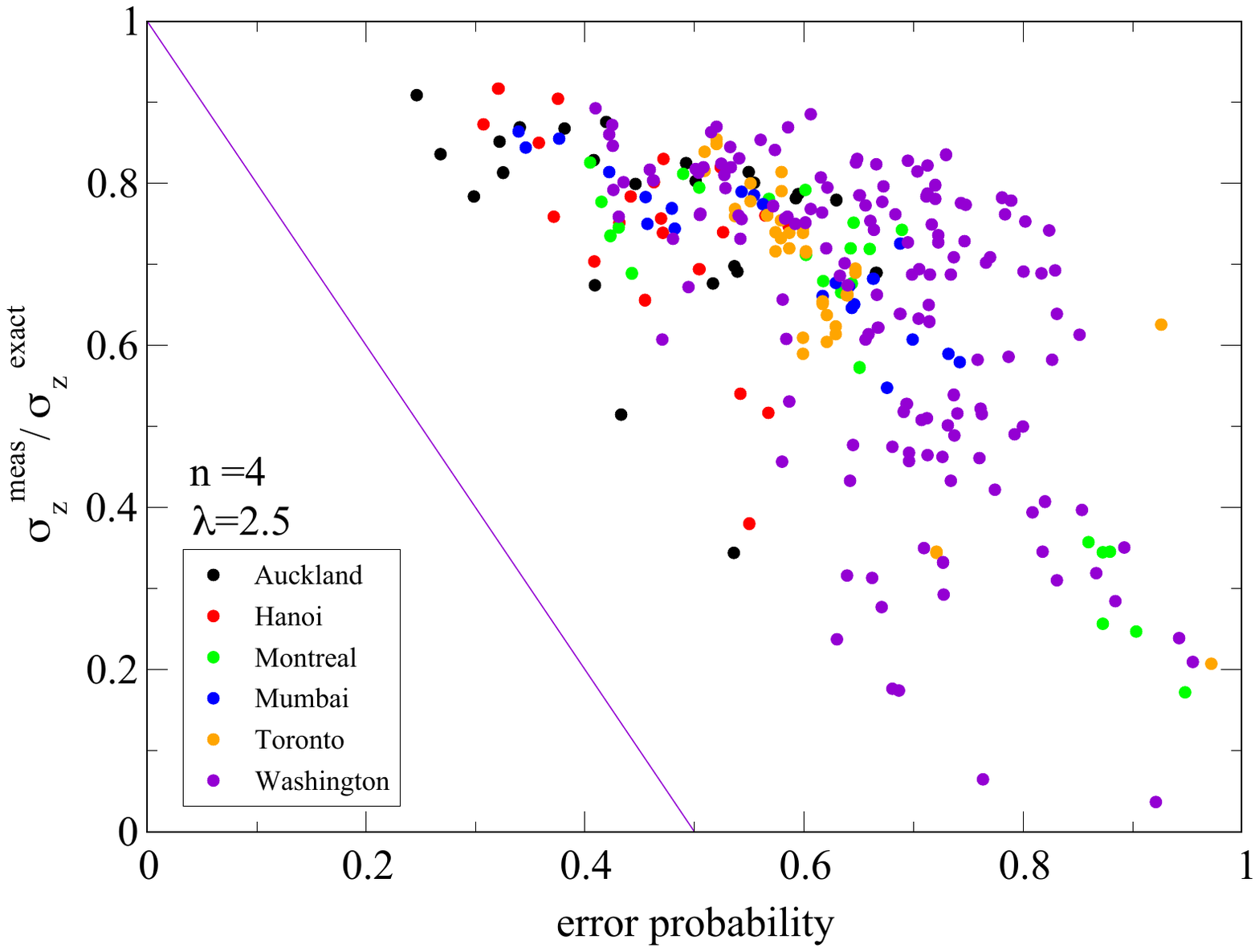}
  \caption{Ratio between the magnetization obtained in the physical qubits in different IBM quantum computers and the exact magnetization for the n=4 Ising model in the large external magnetic field ($\lambda=2.5$) case as a function of the total error probability ($P_{T}$). Each point corresponds with a different qubit chain and each color with a different IBM quantum computer.}
  \label{fig-ising-m}
\end{figure}

The results of the ratio between the measured and the simulated magnetizations for different physical qubit chains are presented in Fig. (\ref{fig-ising-m}). The purple line represents the minimum value of the magnetization if the error provokes the maximum magnetization change due to the total error probability. For this value of $\lambda$, it represents the possibility to switch from a 1 (down) to a 0 (up) for each qubit so the total change can be up to modulus 2. Therefore, the purple line changes linearly from 1 to -1 as the error probability goes from 0 to 1. We can see that all points for all different configurations stand above this line, which indicates that the total error induced by the imperfection of the physical qubits is compatible with the one that is induced taking into account the total error probability we have calculated using our tool. These results can be compared with the ones calculated by Cervera in  Ref. \cite{CerveraIsing2018} which were calculated in 2018 in the IBM quantum computers. The smallest error that we obtain for the calculation of the magnetization is smaller than 7$\%$, while the ones in    Ref. \cite{CerveraIsing2018} for large values of the external magnetic field are in the best scenarios of the 50$\%$. This is an impressive improvement of the performance of the current quantum computers, their error has been reduced more than 7 times in just 4 years. 

\subsection{The quantum phase estimation}
\label{qpe}

The QPE  is an algorithm that permits the calculation of the eigenvalue of any unitary matrix  \cite{Nielsen2011,Dutkiewicz2022} given the eigenstate or eigenvector. The eigenvalues of any unitary matrix $U$ have modulus 1, therefore, its eigenvalue equation can be written as
\begin{equation}
U|\psi>=e^{2\pi i\theta} |\psi>,   
\end{equation}
where $\theta\in\mathbb{R}\hspace{0.1cm}:\hspace{0.1cm}\theta\in[0,1)$.  The QPE algorithm is crucial in quantum computing because all quantum circuits are unitary matrices. Its role is very important in more complex algorithms like Shor's algorithm  \cite{Shor97}. 

We will calculate the QPE for the $\sigma_x$ operator and for the $|+\rangle=(1/\sqrt{2})(|0\rangle+|1\rangle)$ eigenstate, using a total of 4 qubits, one as a register for preparing the eigenstate and other 3 to calculate the phase. Details of the quantum circuit are shown in \ref{appqpe}.

\begin{figure}
  \centering
  \includegraphics[width=0.45\textwidth]{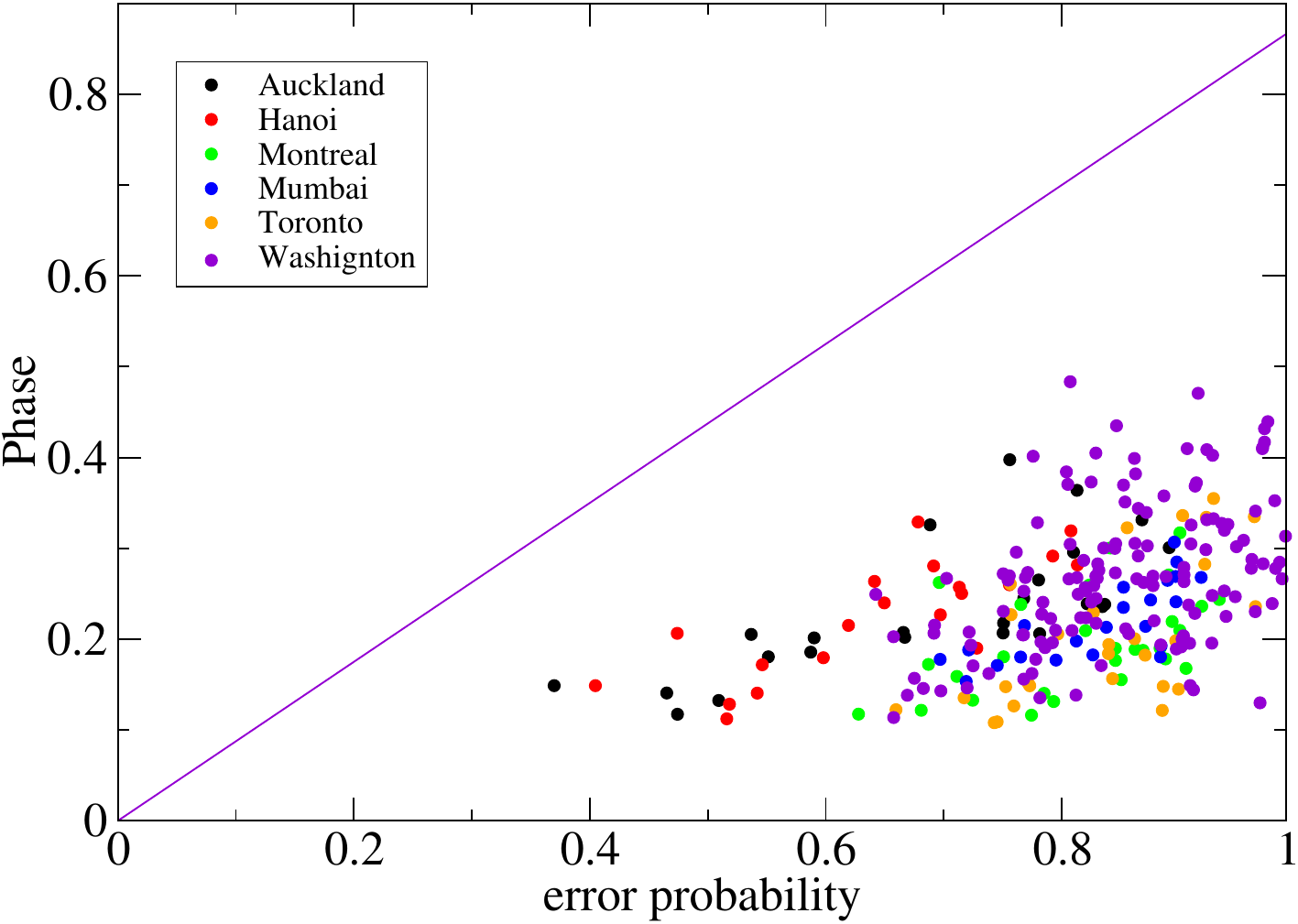}
  \caption{Results of the phase of different qubit chains in different IBM quantum computers for QPE $\sigma_x$ circuit for the $|+\rangle$ state as a function of the total error probability ($P_{T}$). Each point corresponds with a different qubit chain and each color with a different IBM quantum computer.}
  \label{fig-qpe-phase}
\end{figure}

The phase results  for different qubit chains are shown in Fig. (\ref{fig-qpe-phase}). In this case, the expected result for $\theta$ is 0 (the eigenvalue of the $|+\rangle$ state is 1) and the purple line represents the maximum error we can generate as a function of the  error probability. As in the previous case the worst scenario is to obtain $|111\rangle$ instead of $|000\rangle$. For the QPE circuit with 4 qubits this implies that the eigenvalue scales linearly from 0 to 7/8 as a function of the error probability. In this case, all the points stay below this line which indicates that the calculation of the total error probability matches perfectly with the errors induced by the imperfections of the physical qubits.   

\subsection{The Grover's algorithm}
\label{Grover}

The Grover's algorithm, developed by Lov Grover in 1996, is a quantum search algorithm that can greatly improve the efficiency of searching through a large dataset. In classical algorithms, searching for an element that satisfies a certain property typically requires $O(N)$ searches, where N is the size of the dataset. The Grover's algorithm, on the other hand, can perform this search in $O(N^{1/2})$ iterations, making it exactly (and not only asymptotically) optimal \cite{zalka1999grover}.

The Grover's algorithm can be used to find elements that satisfy a wide range of properties, not just simple ones. It can be applied to search for a specific number in a list of numbers, for example. The algorithm will output the target number with a high probability if it is present in the list, and a low probability if it is not. To determine the algorithm's performance, the probability of finding the target element can be used as a measure.

It is also worth noting that the Grover's algorithm can only be used on unstructured databases and it's more efficient than classical algorithms when the number of solutions is smaller than the size of the data set. The algorithm is also known as the quantum search algorithm with quadratic speedup. Details of the quantum circuit are provided in \ref{appgrover}.

\begin{figure}
  \centering
  \includegraphics[width=0.45\textwidth]{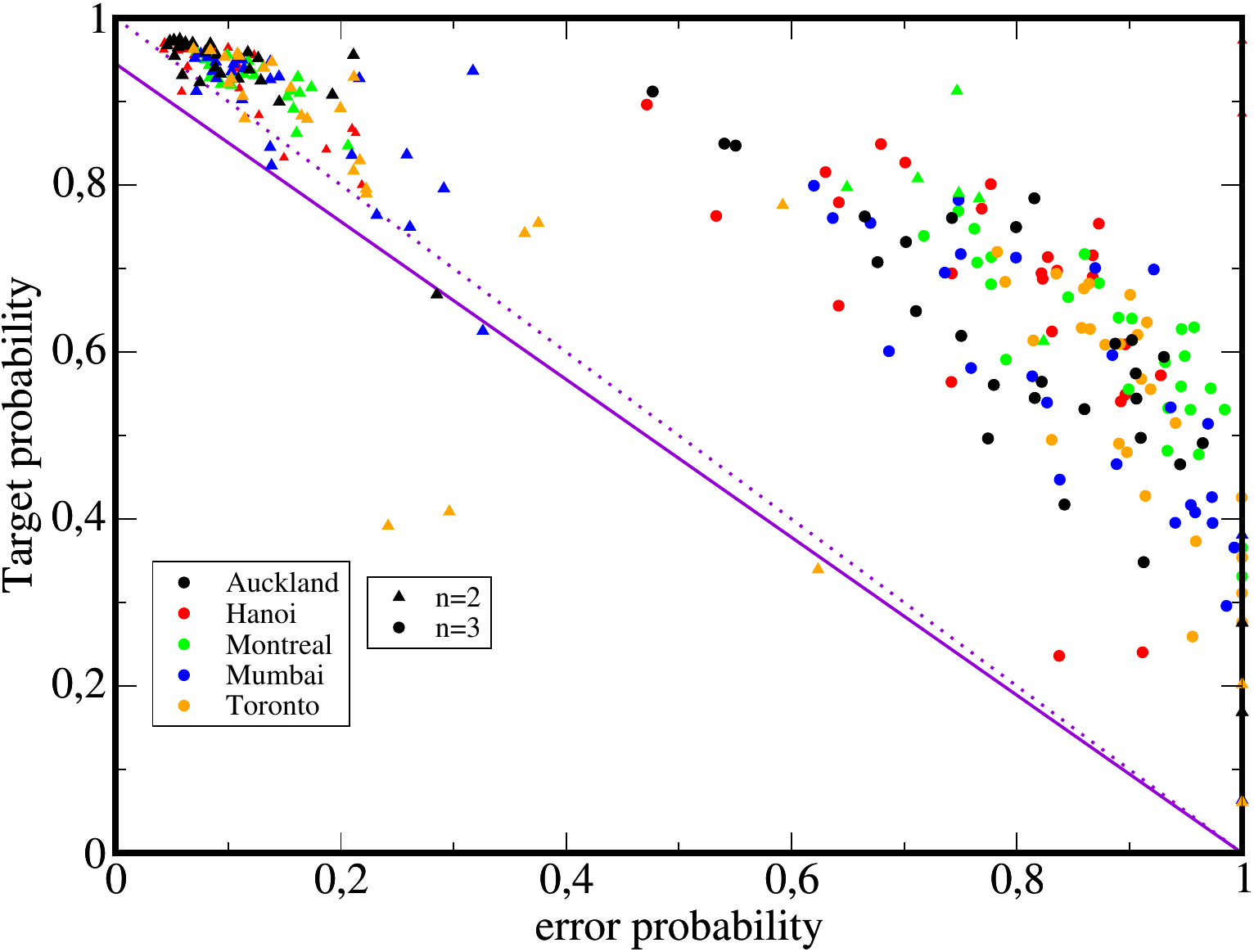}
  \caption{Probability of finding the target state of the Grover's algorithm as a function of the total error probability ($P_{T}$). The total sample size is $2^n=4$ for the triangles and $2^n=8$ for the circles. Each point corresponds with a different qubit chain and each color with a different IBM quantum computer.}
  \label{fig-grov}
\end{figure}

We can see in Fig. (\ref{fig-grov}) the results of the target probability for 2 different sizes of the search list. In the first one, we have $2^n=4$ elements, and we can see that the error probability is, in general very small, leading to high probabilities of finding the target element. However, if we increase the size of our search list up to  $2^n=8$ elements, the total error probability increases dramatically and the probability of finding the target element decreases accordingly. The dashed ($n=2$) and solid (n=3) lines represent the minimum target probability \cite{note_Groover_prob,WatrousQCNotes}.

After evaluating the three different quantum circuits, we are able to conclude that both the information given by the API of IBM and the total  error probability  estimation tool are fully reliable. 

\begin{figure}
  \centering
  \includegraphics[width=0.475\textwidth]{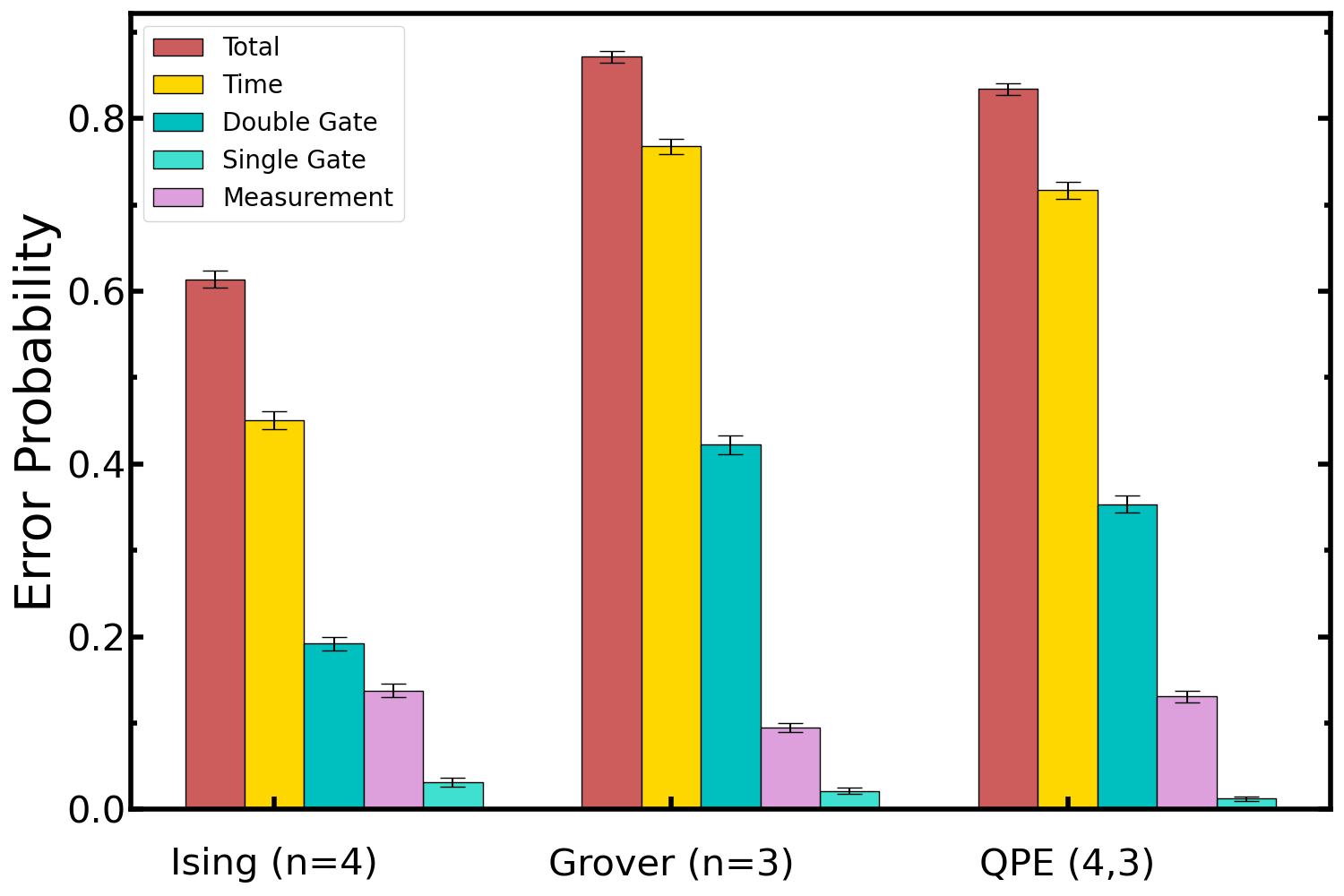}
  \caption{Error contributions in the three different studied quantum circuits. The error contributions are coming from three main sources: Time, measurement, and gate operations (single and double). The error bars correspond to the mean value of all independently derived qubit errors, and the standard deviations are presented.}
  \label{fig-error-bar}
\end{figure}

In order to conclude the section, in Fig. (\ref{fig-error-bar}) we present an error bar plot of our studied circuits highlighting the three main error sources: time-related, measurement-related, and gate operation-related errors, the latter being further differentiated into single and double gate operations. The error bars represent the mean value of all independently derived qubit errors for different qubit chains and quantum computers, with the corresponding standard deviation also presented. In our assessment of the three representative algorithms, the most substantial error source stems from the time factor, with gate operation errors—particularly from double gate operations—being the secondary contributor. Naturally, single-gate operations contribute negligible errors. Lastly, the error magnitude derived from measurement is solely reliant on the considered number of qubits $n$, thereby being independent of the circuit length.

\section{Fidelity}
\label{fidelity_sec}

The fidelity is defined as a measure of similarity between two quantum states. In particular
\begin{equation}
F=\langle\Psi_{sim}|\Psi_{phys}\rangle, 
\end{equation}
where $\Psi_{sim}$ is the state obtained in the simulator and  $\Psi_{phys}$ is the state obtained in the physical qubits. The fidelity calculation is far from trivial. In quantum computing, we do not have access to the full quantum state, but to the probabilities, therefore, methods such as the quantum state tomography \cite{elben2020cross,lanyon2017efficient,flammia2011direct} must be used. Even though the fidelity estimation is an active research field, the methods to calculate it are inefficient and the computation becomes unpractical even for small systems of a few qubits. In this work we consider the exact expression to compute the fidelity \cite{flammia2011direct}
\begin{equation}
\label{fid-formula}
F_{\rho\sigma}=\frac{1}{d}\sum_{k}\langle W_{k}\rangle_{\rho}\langle W_{k}\rangle_{\sigma},
\end{equation}
where $\rho$ is the simulated state and $\sigma$ is the physical state, $d=2^{n}$, being $n$ the number of qubits of the quantum circuit and $k$ has $4^{n}$ values, one for each operator that can be created combining $n$ Pauli matrices. The terms $\langle W_{k}\rangle_{\rho}$ correspond to quantum averages of combinations of Pauli matrices ($W_{k}$) in the $\rho$ state. 
Even if the formula is exact, we can see from Eq. (\ref{fid-formula}) that the number of terms  increases exponentially with the number of qubits. A 4-qubit system, it already contains 256 terms.
As every term is computed in a real quantum machine, the computed fidelity will contain errors.
In order to estimate these errors we may differentiate two things. On the one hand, any single $\langle W_{k}\rangle_{\sigma}$ calculation will carry an error that can be related as follows
\begin{equation}
   \langle W_{k}\rangle_{\sigma}\simeq\langle W_{k}\rangle_{\rho}\pm P_{T\sigma},
\end{equation}
where the total error probability $P_{T\sigma}$ will be considered constant for every $k$. On the other hand, Eq. (\ref{fid-formula}) contains a sum of a high number of circuits ($4^n$). We may assume that all of them are independent, so in order to estimate the total error in the fidelity we can consider the average error and not the maximum error. By doing so the obtained error for the confidence bound (purple line) is exactly $P_{T\sigma}$, and, therefore, by using Eq. (\ref{eq_S}) we can get a value for the success probability, or what is the same, a lower bound for the fidelity (purple line). In this case, the purple line does not represent a strict lower bound because we have considered the average error and not the maximum one.

\begin{figure}
  \centering
  \includegraphics[width=0.45\textwidth]{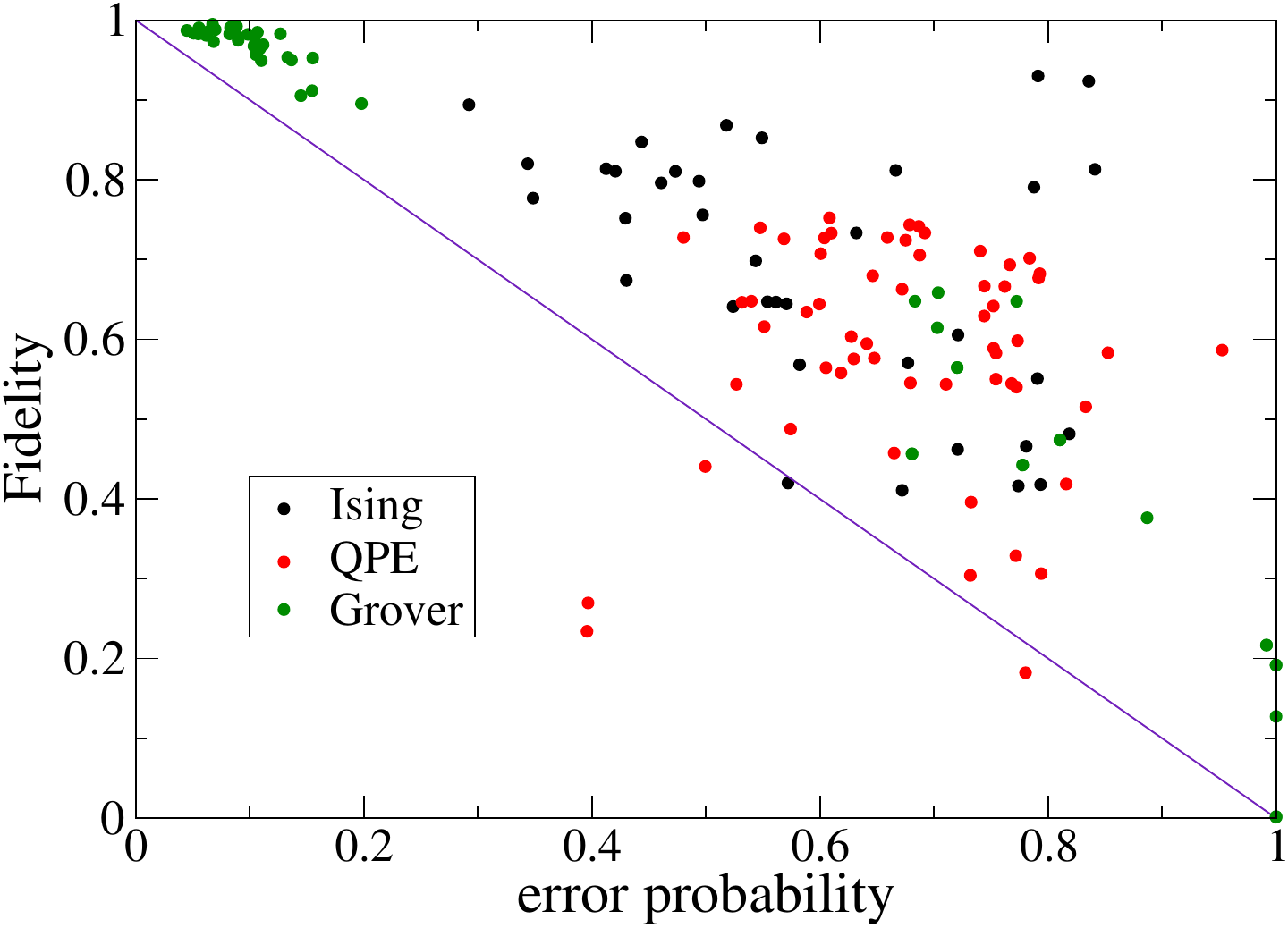}
  \caption{Fidelity as a function of the total error probability ($P_{T}$) for the Ising, QPE, and Grover circuits. The purple line corresponds to the $1-x$ function. The different colors correspond to the different circuits. }
  \label{fidelity}
\end{figure}

The results for the confidence bound (purple line) for the fidelity and the computed values are shown in Fig. (\ref{fidelity}). As we can see our results stay above the confidence line up to the $90\%$ of the cases. 

\section{Measurement error mitigation}
\label{mitigation}

In this section, we evaluate the effect of the error mitigation techniques in correcting the error induced by the measurement. One way of achieving this is by using the Qiskit measurement error mitigation function, which generates a matrix M consisting of measurements over all the basis states of our system.

It's worth noting that the size of this basis grows exponentially as $2^n$, where n is the total number of qubits. This means that for small circuits, the number of jobs is also small, but as the size of the circuit increases, the number of jobs increases dramatically. For example, for $n=10$, it has to perform $1024$ jobs, but for bigger numbers like $n=20$, the number jumps to a million jobs. Due to this, this technique is only suitable for small circuits and its cost can be quite high.



\begin{figure*}
  \centering
  \subfloat{\includegraphics[width=0.335\textwidth]{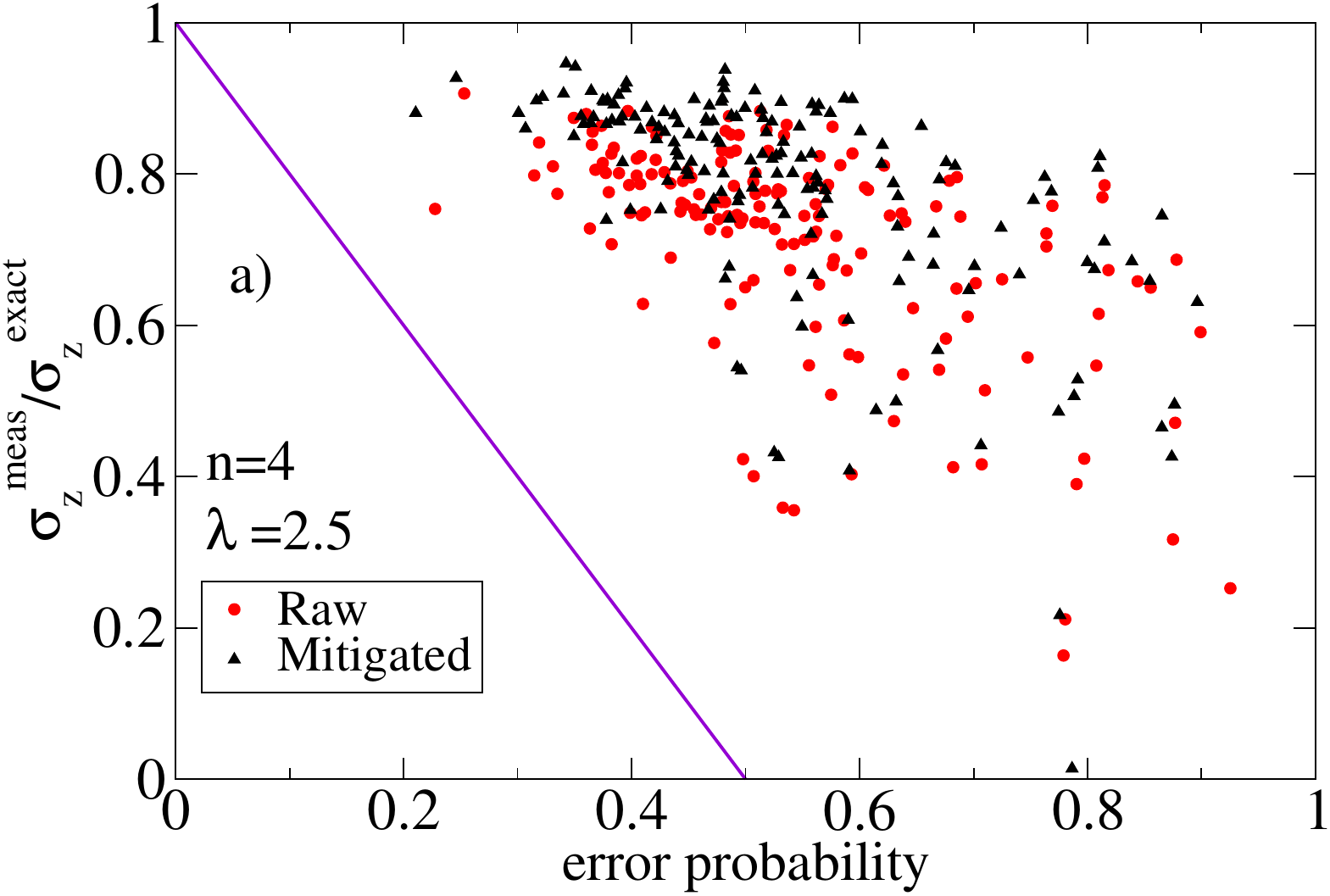}}
  \subfloat{\includegraphics[width=0.31\textwidth]{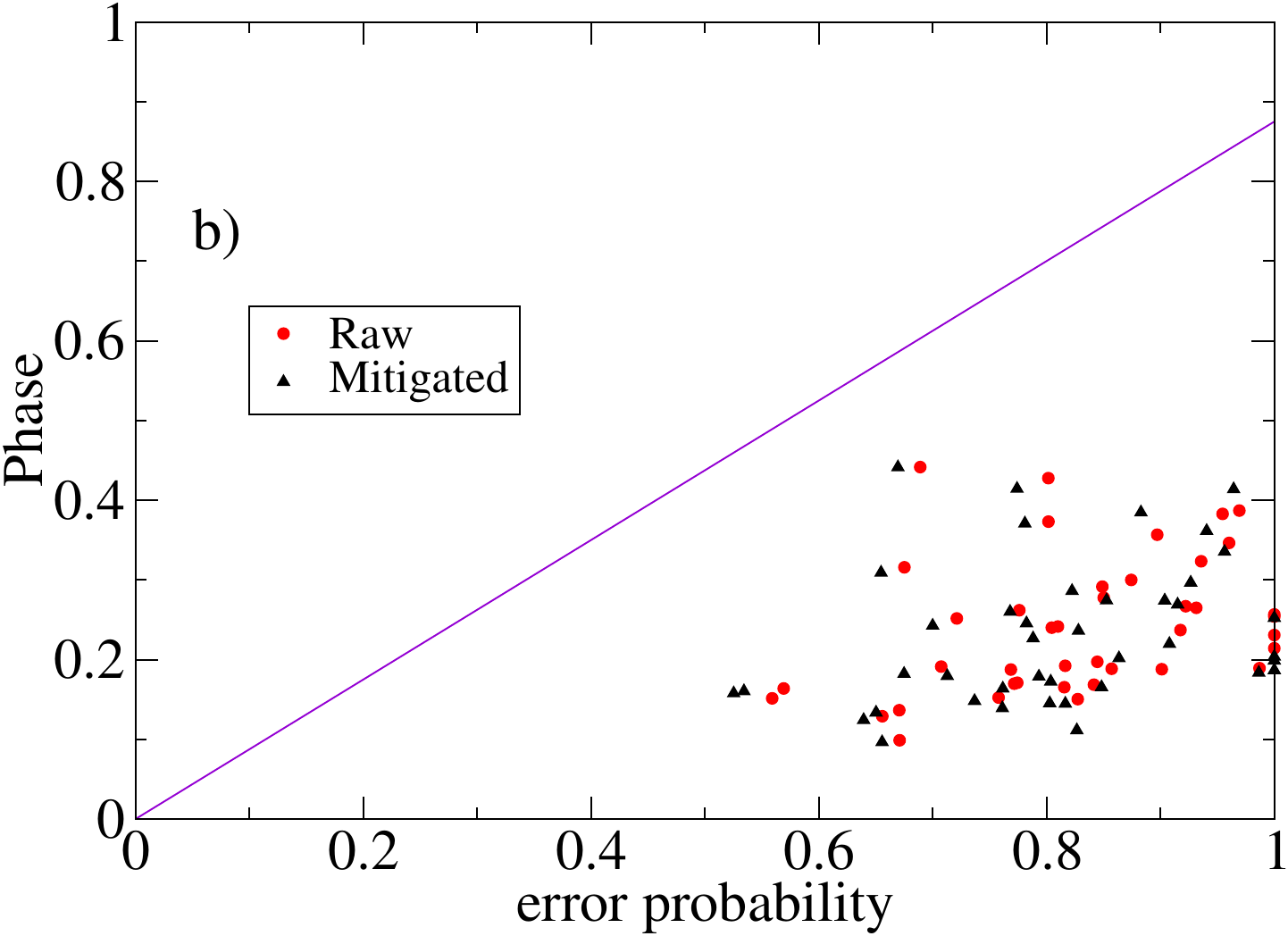}}
  \subfloat{\includegraphics[width=0.315\textwidth]{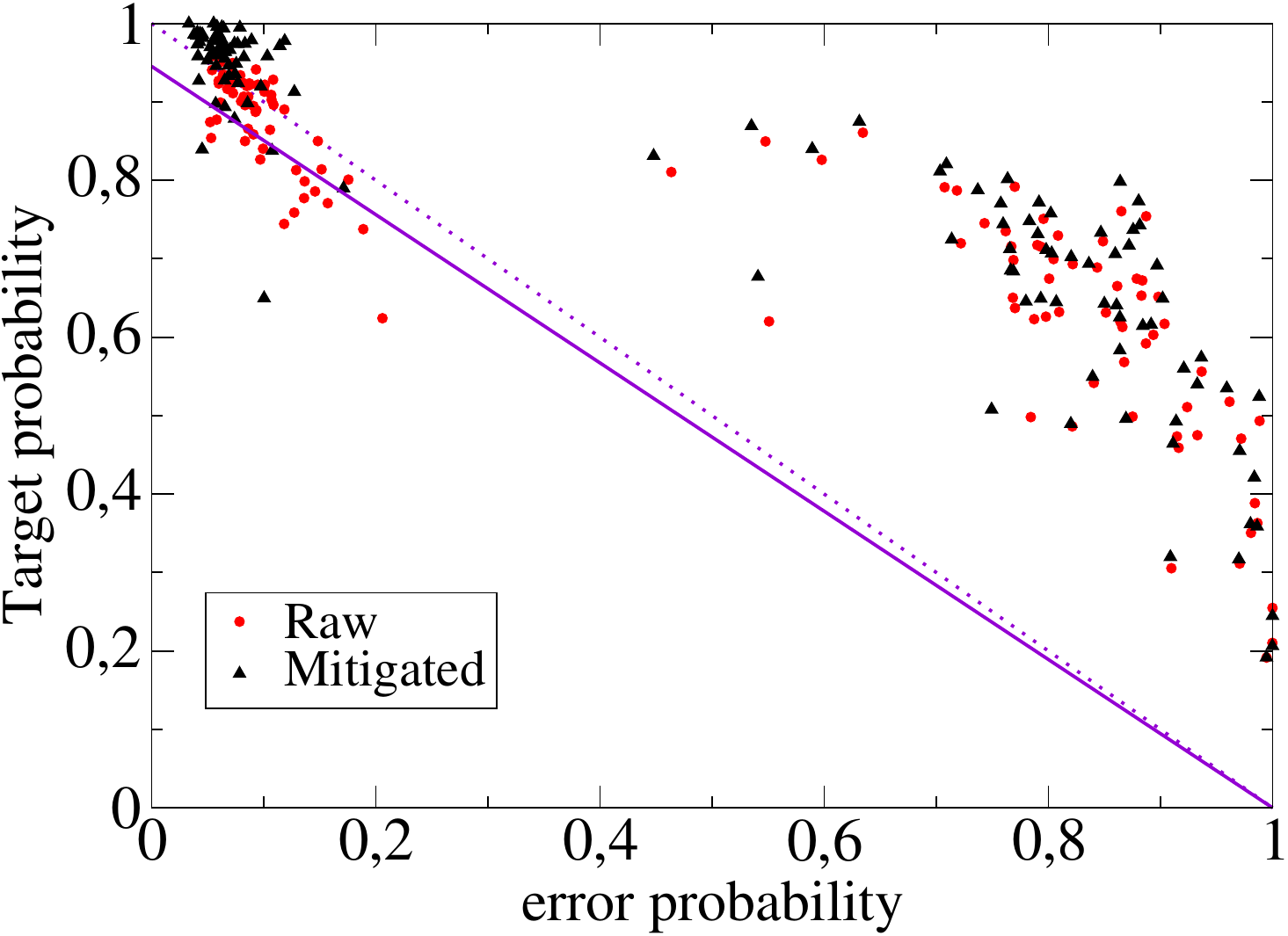}}
  \caption{Magnetization for  Ising model (left), phase of the QPE $\sigma_x$ (center) and Grover's results (right) comparison between the mitigated results (black) and the raw ones as a function of the total error probability ($P_{T}$).}
  \label{fig-mit}
\end{figure*}

To test the effectiveness of this error mitigation routine, we repeated previous calculations using this function. The results are presented in Fig. (\ref{fig-mit}), where we compare the raw and mitigated results for the Ising model, the QPE, and the Grover's algorithm (Fig. (\ref{fig-mit}) a), b), and c), respectively). The raw points have been calculated taking into account the error probability induced by the measurement in Eq. (\ref{St}), while the mitigated ones have been calculated without taking it into account.

The mitigation error technique significantly improves the raw results, as all the points remain above the success probability line (purple line), Fig. (\ref{fig-mit}). This means that the technique is able to eliminate all the errors induced by measurement. However, there are a few points that fall below the purple line, but since they represent less than $1\%$, we can still assume that the mitigation error technique is successful in eliminating measurement errors.

Overall, measurement error mitigation is a crucial aspect of quantum computing and the results presented in this section demonstrate the effectiveness of using the Qiskit measurement error mitigation function. 

\section{Conclusions}
\label{conclusions}
In this work, we have developed a tool (TED-qc) that enables the calculation of the total error probability of any quantum circuit performed in an IBM quantum computer. This is a crucial result in the NISQ era because it permits us to advance the reliability of the result one may obtain in any real quantum computer. 
The algorithm can be run easily on any personal computer, which may help to reduce the unnecessary use of real quantum computers. It is important to remark that it permits us to estimate the error in any quantum calculation without comparing it with the classical one.  Hence, the TED-qc provides a general and extensible framework designed to facilitate further progress in the field. In addition, it can be used as a pre-processing estimator for the lower bound of the fidelity.

In order to prove the robustness of this tool we have realized a big number of different calculations on three representative quantum circuits, the one-dimensional Ising model, the QPE, and the Grover's algorithm. In these cases, we have compared the results of the physical qubits with the ones obtained in the simulator (which is noiseless) and we have printed them as a function of the total error probability. The results are very satisfactory because more than $99\%$  of the errors were smaller than the maximum that could be predicted through the total error probability. Taking into account the statistical nature of the way this  error probability is calculated we can assure that this concept as a measure of the error is both robust and reliable.

We have also studied the effect of the measurement error mitigation routine. This technique eliminates the noise produced during the measurement. In order to do so it has to perform $2^n$ quantum jobs, being $n$ the number of qubits we use in our quantum circuit. We have proven that this technique may be able to eliminate all the errors induced by the measurement. To do so we have studied the results obtained  through the mitigation error as a function of the error probability  which does not include the noise induced by the measurement. We have seen that the mitigated results are compatible with a total error probability  which excludes the noise that occurs during the measurement. Nevertheless, this technique needs a very high number of evaluations and if the number  of qubits we may use is high enough (more than 20) its cost may be too high to be used.

These results have been calculated for the IBM quantum computers and in order to calculate the total error probability we have used the API of the company. Nevertheless, both the tool and the analysis can be easily extended to any other quantum computer following the same lines we have presented here.

\vspace{0.2cm}

\section*{Acknowledgement}
We acknowledge the use of IBM Quantum services for this work. The views expressed are those of the authors and do not reflect the official policy or position of IBM or the IBM Quantum team. 
We thank the support of the Government of Biscay (Bizkaiko Foru Aldundia – Diputación Foral de Bizkaia) through Lantik and its Industry Focused Quantum Ecosystem initiative which provided access to the IBM quantum computers. This work was supported by Programa de Red Guipuzcoana de Ciencia, Tecnología e Innovación
Proyectos de I+D, Convocatoria 2023.

\appendix

\section{Appendix: The quantum circuits}

\subsection{Quantum phase estimation}
\label{appqpe}

The QPE algorithm calculates the phase of the eigenvalue of a unitary matrix $U$ for a proper eigenstate $\psi$.  The QPE uses two registers. The first one is composed of $t$ qubits, the bigger is $t$ the bigger is the precision of the estimation. The second register contains the $\psi$ state. The circuit is described in Fig. (\ref{circuit}) and it consists of applying $t$ Hadamard gates to the $t$ first register qubits and then controlled-$U$ ($U=\sigma_x$ in this case, represented with a $+$ sign in the figure) operations in the way shown in  Fig. (\ref{circuit}). Being $\psi$ (the eigenstate stored in the second register) an eigenstate of $U$ ($\sigma_x$) it will not change during the execution of the quantum circuit as the only gates applied to it are $U$ ($\sigma_{x}$) gates. After the appliance of the H and controlled-$U$ ($\sigma_x$) gates the first register will read
\begin{equation}
\begin{aligned}
\frac{1}{2^{t/2}}\left(|0\rangle+e^{2 \pi i 2^{t-1} \theta}|1\rangle\right)\left(|0\rangle+e^{2 \pi i 2^{t-2} \theta}|1\rangle\right)  \cdots \nonumber\\
\cdots 
\left(|0\rangle+e^{2 \pi i 2^0 \theta}|1\rangle\right).
\end{aligned}
\label{qpe_state}
\end{equation}
If $\theta$ can be expressed exactly by $t$ bits using the binary fraction $\theta=0._{\theta_{1}\dots\theta_{t}}=\frac{\theta_{1}}{2^{1}}+\dots+\frac{\theta_{t}}{2^{t}}\hspace{0.1cm}:\hspace{0.1cm}\theta_{1},\dots,\theta_{t}=0,1$, then Eq. (\ref{qpe_state}) may be rewritten
\begin{equation}
\begin{aligned}
\frac{1}{2^{t/2}}\left(|0\rangle+e^{2 \pi i 0._{\theta_{t}}}|1\rangle\right)\left(|0\rangle+e^{2 \pi i0._{\theta_{t-1}\theta_{t}}}|1\rangle\right)  \cdots \nonumber\\
 \cdots
\left(|0\rangle+e^{2 \pi i0._{\theta_{1}\theta_{2}\dots\theta_{t}}}|1\rangle\right).
\end{aligned}
\label{qpe_state_binary}
\end{equation}
By taking the inverse Fourier transform of Eq. (\ref{qpe_state_binary}) the output is $|\theta_{1}\dots\theta_{t}\rangle$ and, therefore, a measurement in the computational basis will give exactly $\theta$ in its binary fraction form. It can be proven \cite{Nielsen2011}that this method provides a good approximation of $\theta$ even if it cannot be written as a binary fraction of $t$ bits.

\subsection{The Grover's algorithm}
\label{appgrover}

The Grover algorithm is used to solve search problems. Let's understand a search problem as: Given a set $S = \{0, 1,\ldots, 2^{n-1}\}$ of possible solutions, find x belonging to S such that $f(x) = 1$ for a certain function $f$. In addition, we will assume that the element $x$ that satisfies $f(x) = 1$ is unique in $S$, and we will denote it as $w$. Let $n$ be the number of qubits in the circuit.

To solve the search problem, we need to repeatedly apply the Oracle and Amplifier, where the Oracle is a unitary operator $U_w$ that satisfies $$ U_w|x \rangle = (-1)^{f(x)} |x\rangle, \\ \forall x \in S,$$ and the Amplifier is another unitary operator that performs the inversion about the mean of amplitudes. That is, it modifies the amplitude of each state with respect to the mean of all amplitudes.

Therefore, by iterating the Oracle + Amplifier $k$ times, where $k$ is the integer closest to $\frac{\pi}{4}\cdot \sqrt{2^n}$, the probability of not returning the desired element is of $O(1/N)$, being negligible for sufficiently large $N$, with $N = 2^n$.

\begin{figure}
	\centering
	\includegraphics[width=0.8\linewidth]{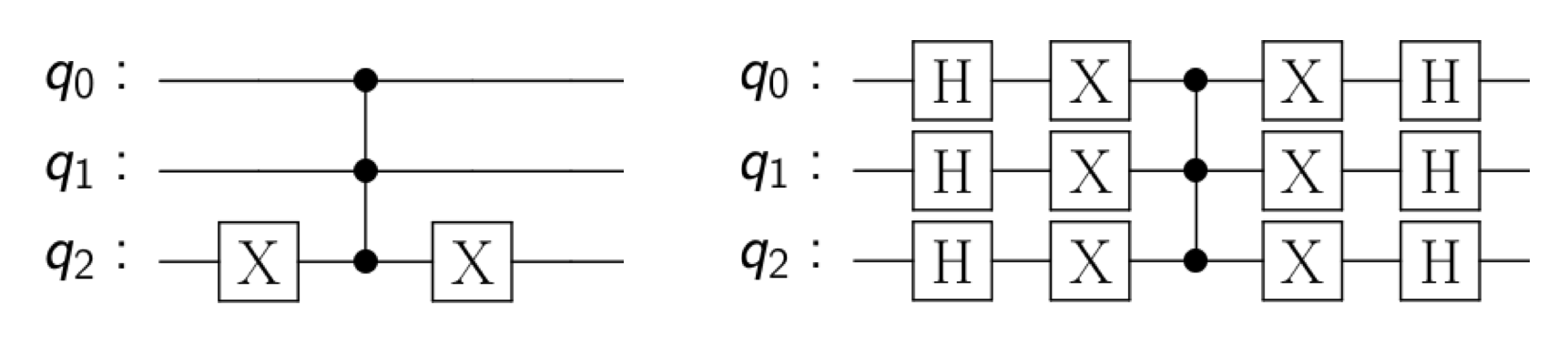}   
	\caption{Oracle (left) and Amplifier (right) implementation in a $n=3$ qubit register and $w = |3\rangle = |011\rangle$. The Oracle transformation is independent of $w$.}
	\label{fig_appendix_Groover_3}
\end{figure}

The construction of the Oracle depends on the element that we are searching $|w\rangle = |b_{n-1} .. b_1 b_0\rangle$. First, the Oracle applies an $X$ gate to the $i-th$ qubit if the element $b_i=0$, for all $i=0,\ldots,n-1$. Then a multi-control-Z gate is applied on all qubits, and finally, an $X$ gate is applied to the $i-th$ qubit if the element $b_i=0$. See Fig.~\ref{fig_appendix_Groover_3} (left)  for the Oracle's implementation in the particular case for $n = 3$ qubits and $w = |3\rangle = |011\rangle$.

The Amplifier is constructed by applying a column of Hadamard gates on all qubits, a column of $X$ gates on all qubits, a multi-control-Z gate on all qubits, and symmetrically a column of X gates followed by a column of Hadamard gates on all qubits, as can be appreciated in  Fig.~\ref{fig_appendix_Groover_3} (right) for $n=3$.

\section{Declarations}
The authors have no relevant financial or non-financial interests to disclose.

\bibliography{refs_SJ-3 }

\begin{thebibliography}{50}
\bibitem [{not()}]{note_Groover_prob}%
  \BibitemOpen
  \href@noop {} {}\bibinfo {note} {In a noisless evaluation of the Grover's
  algorithm, the probability of finding the target state after k iterations is
  $P = \sin^2\left((2k+1)\theta\right)$ with $\theta=\arcsin(2^{-n/2})$. Since
  we consider $k\approx\frac{\pi}{4}\sqrt{2^{n}}-\frac{1}{2}$ we can write $P =
  \sin^2\left(\left(\pi2^{\frac{n}{2}-1}+1\right)\arcsin\left(2^{-\frac{n}{2}}\right)\right)$}\BibitemShut
  {NoStop}%
\end{thebibliography}


\end{document}